\documentstyle[12pt]{article}
\catcode`\@=11
\def\maketitle{\par
 \begingroup
 \def\thefootnote{\fnsymbol{footnote}}
 \def\@makefnmark{\hbox
 to 0pt{$^{\@thefnmark}$\hss}}
 \if@twocolumn
 \twocolumn[\@maketitle]
 \else \newpage
 \global\@topnum\z@ \@maketitle \fi\thispagestyle{empty}\@thanks
 \endgroup
 \setcounter{footnote}{0}
 \let\maketitle\relax
 \let\@maketitle\relax
 \gdef\@thanks{}\gdef\@author{}\gdef\@title{}\let\thanks\relax}
\def\@maketitle{\newpage
 \null
 \hbox to\textwidth{\hfil\hbox{\begin{tabular}{r}\@preprint\end{tabular}}}
 \vskip 2em \begin{center}
 {\Large\bf \@title \par} \vskip 1.5em {\normalsize \lineskip .5em
\begin{tabular}[t]{c}\@author
 \end{tabular}\par}
 \end{center}
 \par
 \vskip 1.5em}
\def\preprint#1{\gdef\@preprint{#1}}
\def\abstract{\if@twocolumn
\section*{Abstract}
\else \normalsize
\begin{center}
{\large\bf Abstract\vspace{-.5em}\vspace{0pt}}
\end{center}
\quotation
\fi}
\def\endabstract{\if@twocolumn\else\endquotation\fi}
\catcode`\@=12

\begin{document}
\baselineskip=.29in

\preprint{SNUTP-97-020\\[-2mm] hep-th/9703193}

\title{\Large {\bf Self-dual Chern-Simons Solitons\\
in the Planar Ferromagnet}
\protect\\[1mm]\  }
\author{\normalsize Yoonbai Kim${}^{1}$, Phillial Oh${}^{1}$ and Chaiho
Rim${}^{2}$\\
{\normalsize\it ${}^{1}$Department of Physics, Sung Kyun Kwan University,
Suwon 440-746, Korea}\\
{\normalsize\it yoonbai$@$cosmos.skku.ac.kr, ploh$@$newton.skku.ac.kr}\\
{\normalsize\it ${}^{2}$Department of Physics, Chonbuk National University,
Chonju 561-756, Korea}\\
{\normalsize\it rim$@$phy0.chonbuk.ac.kr} 
}
\date{}
\maketitle

\begin{center}
{\large\bf Abstract}\\[3mm]
\end{center}
\indent\indent We consider a uniaxial planar ferromagnet coupled minimally to 
an Abelian Chern-Simons gauge field and study self-dual solitons which 
saturate the Bogomol'nyi bound. We find a rich structure of rotationally
symmetric static soliton solutions for various uniform background charge 
densities. For a given ferromagnet material, the properties of these solitons 
are controlled only by the external magnetic field and the background charge.

\vspace{1.5mm}
\noindent PACS number(s): 03.65.Ge, 11.27.+d, 11.10.Lm

\newpage

\pagenumbering{arabic}
\thispagestyle{plain}

The magnetic solitons of the uniaxial ferromagnetic crystals have attracted
much attention recently \cite{KIK}. It is well known that the 
macroscopic properties of a ferromagnet material are described by the 
Landau-Lifshitz equation (LLE) \cite{LL} and the order parameters are the spin 
variables $Q\equiv Q^{a}(t,x^{i})T^{a},\;T^{a}=-i\sigma^{a}/2$ :
\begin{equation}\label{ll}
\partial_{t}Q+\partial_{i}[Q,\partial_{i}Q]-\sum_{a=1}^{3}
(J^aQ^{a}-H^a)[Q,T^{a}]=0,
\hspace{15mm} Q^{a}Q^{a}=1.
\end{equation}
The above equation is derived by assuming the potential energy of the form
up to a constant
\begin{equation}\label{aisoH}
W_a=\sum_{a=1}^{3}(\frac{1}{2}J^aQ^{a}Q^a-H^a Q^a).
\end{equation}
The first term of the potential energy accounts for the magnetic anisotropy 
energy and, for a uniaxial system, $J^{a}=\lambda\delta_{3a}$. 
If $\lambda<0$, it describes a ferromagnet of the easy axis type and, if 
$\lambda>0$, the easy plane type. The second term depicts
the external magnetic energy and we will assume from here on 
$H^a=H\delta_{a3}$.  

In (1+1) dimensions, LLE supports domain walls. In (2+1) dimensions, 
prototypical solitons are Belavin-Polyakov type lump solutions, which exist 
when there is no anisotropy energy \cite{BP}. It is obvious by the Derrick's
theorem that these two are the whole spectra of static solitons with finite 
energy supported by LLE. Furthermore, the Belavin-Polyakov lump solutions are 
conformally invariant, which makes them less applicable to realistic cases. 
Phenomenologically, there exist rich spectra of solitons without a scale 
invariance \cite{KIK}. To understand these magnetic solitons theoretically, one
may include the higher-order spatial derivative terms like Skyrmions
\cite{ISZ}, or consider charged objects as stationary solutions \cite{Lee}. 

Here let us recall the relation between the superfluidity with a {\it global}
$U(1)$ symmetry and the superconductivity of the Landau-Ginzburg theory of 
a {\it local} $U(1)$ symmetry. In superfluidity, there are global vortices 
with logarithmically divergent energy, whereas, in superconductivity, local
vortices have finite energy. Thus gauging a theory  provides a useful 
mechanism for obtaining the finite energy solitons. 
Among these local vortices, those 
which saturate Bogomol'nyi limit are important ones because they give a 
criterion to distinguish between type-I and type-II superconductivity in 
Maxwell theory \cite{Bog}. It turns out that the Bogomol'nyi limit plays also
a crucial role in finding a variety of self-dual solitons of intriguing
properties in the Chern-Simons (CS) gauge theories \cite{HKP,JP}. Recent 
developments along 
these directions have been in the study of self-dual solitons in the gauged 
relativistic $O(3)$ nonlinear sigma model with the Maxwell term \cite{Sch},
the non-Abelian or the Abelian Chern-Simons term \cite{Nar,GG}. In these 
models, the $O(3)$ global symmetry is broken down to $U(1)$ without a clear
physical motivation. On the other hand, in the nonrelativistic uniaxial 
ferromagnet system described by Eq.(\ref{ll}), the $O(3)$ global symmetry is 
broken explicitly to $U(1)$ by the relativistic effect \cite{LL}. Therefore, 
it is natural to couple an Abelian gauge field to such $U(1)$ direction in the 
nonrelativistic theory. 

Recently, a general way of gauging the ferromagnet defined on the Hermitian
symmetric space $G/H$ was proposed, in which the maximal torus subgroup of $H$
was gauged and self-dual equations were obtained \cite{op}. In this Letter, we 
will investigate the gauged LLE of the uniaxial ferromagnet system coupled to
CS gauge field and study the self-dual solitons which saturate Bogomol'nyi 
limit.

We start from the action which is a $SU(2)$ version of the
generalized planar anisotropic Chern-Simons ferromagnet 
coupled with the uniform background charge:
\begin{equation}\label{action}
S=\int dt\,d^{2}x\biggl
\{\mbox{Tr}\Bigl[2Kg^{-1}D_{t}g+D_{i}QD_{i}Q \Bigr]-\rho_eA_0-W_a(Q)
\biggr\}+S_{g}(A_{\mu}).
\end{equation}
Here, $g\in SU(2)$, $K=i\,\mbox{diag}(1/2,-1/2)$, $Q=gKg^{-1}$, and the
covariant derivative is given by
\begin{equation}
D_{\mu}g=\partial_{\mu}g+A_{\mu}T^{3}g,\hspace{10mm}D_{i}Q=\partial_{i}Q+A_{i}
[T^{3},Q].
\end{equation}
$\rho_{e}$ is the uniform  charge density 
which will be responsible for a rich structure of self-dual solitons.
The magnetic anisotropy energy $W_a$ in Eq.(\ref{aisoH}) is given by
\begin{equation}\label{magn}
W_a=\frac{\lambda}{2}(Q^3-v)(Q^3+\rho_e).
\end{equation}
For the gauge dynamics, we consider the Abelian  Chern-Simons term
\begin{equation}
S_{g}(A_{\mu})=\frac{\kappa}{2}\epsilon^{\mu\nu\rho}\int dt\,d^{2}x
A_{\mu}\partial_{\nu}A_{\rho}.
\end{equation} 
The above action is invariant under a left CS gauge transformation:
\begin{eqnarray}
 g&\rightarrow& hg,~ h=\mbox{diag}(\exp(-i\Lambda/2),\exp(i\Lambda/2))
 \nonumber\\
Q&\rightarrow& hQh^{-1},~ A_\mu\rightarrow A_\mu+\partial_\mu \Lambda.
\end{eqnarray} 
Note that it is also invariant under a right local transformation, 
$g\rightarrow gh$, which corresponds to the $U(1)$ symmetry of the ungauged 
$CP(1)$ model \cite{op1}. 

The Euler-Lagrange equation in terms of spin variables $Q$ becomes the gauged 
inhomogeneous LLE:
\begin{equation}
D_tQ+D_i[Q,D_iQ]-\lambda(Q^3-\frac{v-\rho_{e}}{2}) [Q,T^{a}]=0.
\end{equation}
The gauge field is governed by the Chern-Simons equation:
\begin{equation} 
\frac{\kappa}{2}\epsilon^{\mu\nu\rho}F_{\nu\rho}=j^{\mu}.
\end{equation}
Here $j^{\mu}$ is a conserved current expressed by
\begin{equation}
j^{\mu}\equiv(\rho,j^{i})=(Q^{3}+\rho_e,\;2\mbox{Tr}K[Q,D_{i}Q]),
\end{equation}
and then $Q_{{\rm U(1)}}=\int d^{2}x Q^3$ is conserved $U(1)$ charge.

The gauge invariant topological current is given by
\begin{eqnarray}
T^{\mu}&=&\frac{1}{8\pi}\epsilon^{\mu\nu\rho}\Bigl[\epsilon^{abc}Q^{a}
D_{\nu}Q^{b}D_{\rho}Q^{c}+F_{\nu\rho}(w-Q^{3})\Bigr]\label{topcu3}\\
&=&\frac{1}{8\pi}\epsilon^{\mu\nu\rho}\epsilon^{abc}Q^{a}\partial_{\nu}
Q^{b}\partial_{\rho}Q^{c}+\frac{1}{4\pi}\epsilon^{\mu\nu\rho}\partial_{\nu}
\bigl((w-Q^{3})A_{\rho}\bigr),\label{topcu2}
\end{eqnarray}
where $w$ is a free parameter. Note that this topological charge is reduced to 
the winding between a two sphere of compactified 2D space and that of the
configuration space of $Q^{a}$ when the gauge field vanishes.  Eq. 
(\ref{topcu2}) tells us that the gauged topological current differs from the 
ungauged one by the curl of a vector field, and therefore the magnetic flux
also contributes to the conserved topological charge $T\equiv\int
d^{2}x~T^{0}$. 

A kinematical quantity specifying the CS solitons as the anyons is the angular 
momentum and its gauge field contribution is defined by
\begin{equation}\label{angm}
J=\int d^2x \epsilon_{ij}x_i\Bigl\{ 2\mbox{Tr}[ -T^{3}(g^{-1}\partial_{j}g
-(g^{-1}\partial_{j}g)|_{|\vec{x}|\rightarrow \infty})]-A_jQ^3\Bigr\},
\end{equation}
where $|\vec{x}|\rightarrow\infty$ denotes the boundary value of the soliton 
$g(\infty)$ for $\rho_{e}=0$, and the momentum density $p_i$ is given by
\begin{equation}
p_i=2\mbox{Tr}(-T^{3}g^{-1}D_{i} g).
\end{equation}
Here we count only the unambiguous contribution of the angular momentum from 
the  CS gauge field, which is finite in the limit of zero background CS charge
\cite{PT}.

Let us derive the Bogomol'nyi limit of the Chern-Simons ferromagnet system 
in Eq. (\ref{action}) under the specific anisotropic energy in Eq. 
(\ref{magn}). When $\lambda=\mp\frac{2}{\kappa}$, the Chern-Simons system 
achieves the Bogomol'nyi bound:
\begin{eqnarray}\label{csbogh}
E&=&\int d^{2}x\biggl\{\partial_{i}\Bigl(-\frac{\kappa}{2}\epsilon_{ij}
A_0A_{j}\Bigr)+A_0(\kappa B-Q^3-\rho_e)+\frac{1}{2}(D_{i}Q^a)^2+W_a\biggl\}
\nonumber\\
&=&\int d^{2}x \frac{1}{4}|D_{i}Q^{a}\pm\epsilon^{abc}\epsilon_{ij}
Q^{b}D_{j}Q^{c}|^{2}\pm 4\pi T,
\end{eqnarray}
where $T$ is the topological charge in which the free parameter $w$ is fixed
to be the parameter $v$ in Eq. (\ref{magn}), and we used the Gauss' law 
($B\equiv \frac{1}{2}\epsilon_{ij}F_{ij}$)
\begin{equation}
\kappa B-Q^3-\rho_e=0.
\label{gausslaw}
\end{equation}
Note that the spatial integration of the above Gauss' law neglecting the 
uniform charge density term $\rho_{e}$ tells us that any flux-carrying 
Chern-Simons solitons are charged in this model;
$\Phi\Bigl(\equiv\int d^{2}xB\Bigr)=\frac{Q_{U(1)}}{\kappa}$. 

The self-dual solitons which saturate the Bogomol'nyi bound satisfy the
self-dual equations:
\begin{equation}
D_iQ^a=\mp \epsilon^{abc}\epsilon_{ij}Q^bD_jQ^c.
\label{bogeq}
\end{equation}
Introducing a parameterization of the spherical coordinates 
$Q^{a}=(\sin F\cos\Theta,\sin F\sin\Theta,$ $\cos F)$, we express the gauge 
field $A_{i}$ in terms of the scalar fields by solving Eq. (\ref{bogeq}):
\begin{equation}\label{ai}
A_{i}=-\partial_{i}\Theta\mp\epsilon_{ij}\partial_{j}\ln\tan\frac{F}{2}.
\end{equation}
Substituting Eq. (\ref{ai}) into the Gauss law in Eq. (\ref{gausslaw}), we 
obtain a scalar equation for the soliton configurations:
\begin{eqnarray}\label{phieq}
\nabla^{2}\phi\mp\epsilon^{ij}\partial_{i}\partial_{j}\Theta=-\frac{dV}{d\phi},
\end{eqnarray}
where $\phi=\ln\tan\frac{F}{2}$. The shapes of ``effective'' potential for
scalar field $\phi$ are (See Fig. 1)
\begin{equation}\label{effp}
V(\phi)=\pm\frac{1}{\kappa}(\ln\cosh\phi-\rho_e\phi).
\end{equation}



\setlength{\unitlength}{0.1bp}
\begin{picture}(3600,2160)(0,0)
\put(2008,-51){\makebox(0,0){\large $\phi$}}
\put(3778,570){\makebox(0,0){$(-1,-0.5)$}}
\put(3698,840){\makebox(0,0){$(-1,0)$}}
\put(3648,950){\makebox(0,0){$(1,1)$}}
\put(3698,1410){\makebox(0,0){$(-1,1)$}}
\put(3698,2050){\makebox(0,0){$(1,-1)$}}
\put(360,1180){%
\makebox(0,0)[b]{\shortstack{\large $V(\phi)$}}%
}
\put(3417,151){\makebox(0,0){1.5}}
\put(2948,151){\makebox(0,0){1}}
\put(2478,151){\makebox(0,0){0.5}}
\put(2009,151){\makebox(0,0){0}}
\put(1539,151){\makebox(0,0){-0.5}}
\put(1070,151){\makebox(0,0){-1}}
\put(600,151){\makebox(0,0){-1.5}}
\put(540,1923){\makebox(0,0)[r]{2}}
\put(540,1552){\makebox(0,0)[r]{1}}
\put(540,1180){\makebox(0,0)[r]{0}}
\put(540,808){\makebox(0,0)[r]{-1}}
\put(540,437){\makebox(0,0)[r]{-2}}
\end{picture}

\vspace{5mm}

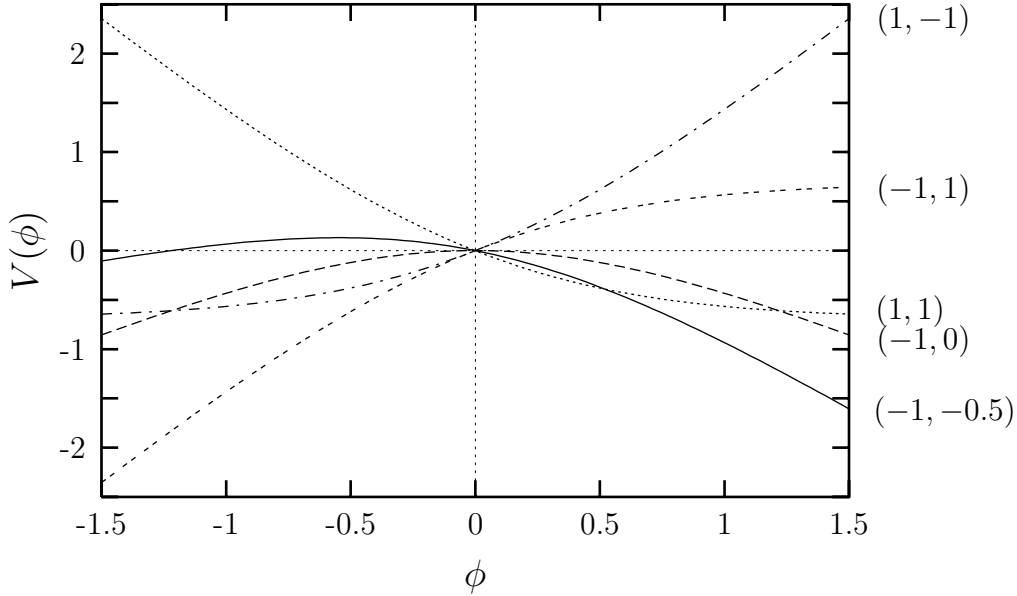
\noindent Figure 1. Shapes of $V(\phi)$ for the sign of $\kappa$ and various
values of $\rho_{e}$, {\it e.g.}, $(-1,1)$ means negative $\kappa$ and 
$\rho_{e}=1$.

\vspace{5mm}

\noindent One can easily notice that the Bogomol'nyi equations in Eq. 
(\ref{phieq}) and Eq. (\ref{effp}) are independent of the external magnetic 
field $H$ in contrast to the fact that the vacuum configuration of the spin 
variable and the dispersion relation of the magnon are independent of the 
background CS charge.

Let us concentrate on the upper sign (self-dual) in Eq.(\ref{bogeq})$\sim$
Eq.(\ref{effp}). The anti self-dual case can be reached by replacing $\kappa$
with $-\kappa$.
For the rotationally symmetric solutions, the ansatz in the cylindrical 
coordinate  $(r,\theta)$ is given  by 
\begin{eqnarray}\label{ansat}
\phi=\phi(r),\;\Theta=n\theta, \;
A_i=\frac{\epsilon_{ij}x_j}{r^2}a(r).
\end{eqnarray}
Then the equation of motion in Eq. (\ref{phieq}) becomes an analogue of the one
dimensional Newton's equation for $r>0$, if we regard $r$ as ``time'' and 
$\phi$ as the position of the hypothetical particle with unit mass:
\begin{equation}\label{newtoneq}
\frac{d^{2}\phi}{dr^{2}}=-\frac{dV(\phi)}{d\phi}-\frac{1}{r}\frac{d\phi}{dr}.
\end{equation}
The exerting forces are the conservative force from the effective potential
$V(\phi)$ in Eq. (\ref{effp}) and time-dependent friction
$-\frac{1}{r}\frac{d\phi}{dr}$. When $n\ne 0$, there is an impact term at $r=0$
due to $\epsilon^{ij}\partial_{i}\partial_{j}n\theta=\frac{n}{r}\delta(r)$
in Eq. (\ref{phieq}).

The boundary conditions for the regular soliton configurations are the
followings. (i) At the origin, $nF(0)=0$ for the singlevaluedness of $Q^{a}$, 
and $a(0)=0$ for the gauge field $A_{i}$ to be non-singular. (ii) At the 
spatial infinity, $F(\infty)$ is determined by the condition 
$\frac{dV(\phi)}{d\phi}\Bigr|_{\phi(\infty)}=0$, and the finiteness of energy
requires $a(\infty)=n$ when $F(\infty)$ is neither zero nor $\pi$. 

Under the ansatz in Eq. (\ref{ansat}), the topological charge $T$ is expressed
by
\begin{equation}
T=\frac{n}{2}\left[\cos F(0)-\cos F(\infty)\right]
+\frac{a(\infty)}{2}(\cos F(\infty)-\rho_{e}+\kappa H),
\end{equation}
and the magnetic flux $\Phi$ (or equivalently the $U(1)$ charge) is given by
\begin{equation}
\Phi=-2\pi a(\infty).
\end{equation}
A straight forward computation of Eq. (\ref{angm}) in terms of Eq. (\ref{ansat})
yields the angular momentum when $\rho_{e}=0$:
\begin{equation}
J|_{\rho_{e}=0}=-\pi\kappa a(\infty)(a(\infty)-2n).
\end{equation}

\vspace{8mm}

\begin{tabular}{|c|c|c|c|c|} \hline
$(\kappa,\rho_{e})$ & $(-,1)$ & $(-,-1<\rho_{e}<1)$ & $(+,+1),\,(1,-1)$ & 
$(+,-1)$ \\ \hline\hline
$F(0)$ & 0 & 0 & $0<F(0)<\infty$ & 0 \\ \hline
$F(\infty)$ & $\pi$ & $2\tan^{-1}\sqrt{\frac{1-\rho_{e}}{1+\rho_{e}}}$ & 0
& 0\\ \hline
$a(\infty)$ & $\alpha,\,(0<\alpha<n-1)$ & $n$ & $\alpha,\,(\rho_{e}\alpha<-1)$ 
& $\alpha,\,(\alpha>n+1)$\\ \hline
$T$ & $n-\alpha(1-\frac{\kappa H}{2})$ &$\frac{n}{2}(1-\rho_{e}+\kappa H)$ 
& $\frac{\alpha}{2}(1-\rho_{e}+\kappa H)$&
$\frac{\alpha}{2}(2+\kappa H)$ \\ \hline
$\Phi$ & $-2\pi\alpha$ & $-2\pi n$ & $-2\pi \alpha$ & $-2\pi \alpha$\\ \hline
$J|_{\rho_{e}=0}$ & $-\pi\kappa\alpha(\alpha-2n)$ &$\pi\kappa n^{2}$ &
$-\pi\kappa\alpha^{2}$ &
$-\pi\kappa\alpha(\alpha-2n)$\\ \hline
Species & topological & topological & nontopological & nontopological \\
& lump $(n\geq 2)$ & vortex & soliton & vortex \\ \hline
\end{tabular}

\vspace{6mm}

\noindent Table 2. Self-dual CS solitons for the sign of $\kappa$ and various 
values of $\rho_{e}$ ($n$ means a positive integer and $\alpha$ means an 
arbitrary real number).



\setlength{\unitlength}{0.1bp}
\begin{picture}(3600,2160)(0,0)
\put(2008,-51){\makebox(0,0){\large $r$}}
\put(3868,1950){\makebox(0,0){$(-1,1),\,n=2$}}
\put(3818,1800){\makebox(0,0){$(1,1),\,n=0$}}
\put(3858,1090){\makebox(0,0){$(-1,0),\,n=1$}}
\put(3948,800){\makebox(0,0){$(-1,-0.5),\,n=1$}}
\put(3918,280){\makebox(0,0){$(1,-1),\,n=0,1$}}
\put(300,1180){%
\makebox(0,0)[b]{\shortstack{\large $F(r)$}}%
}
\put(3417,151){\makebox(0,0){8}}
\put(3065,151){\makebox(0,0){7}}
\put(2713,151){\makebox(0,0){6}}
\put(2361,151){\makebox(0,0){5}}
\put(2009,151){\makebox(0,0){4}}
\put(1656,151){\makebox(0,0){3}}
\put(1304,151){\makebox(0,0){2}}
\put(952,151){\makebox(0,0){1}}
\put(600,151){\makebox(0,0){0}}
\put(540,1844){\makebox(0,0)[r]{3}}
\put(540,1313){\makebox(0,0)[r]{2}}
\put(540,782){\makebox(0,0)[r]{1}}
\put(540,251){\makebox(0,0)[r]{0}}
\end{picture}

\vspace{15mm}


\setlength{\unitlength}{0.1bp}
\begin{picture}(3600,2160)(0,0)
\put(2008,-51){\makebox(0,0){\large $r$}}
\put(3848,2010){\makebox(0,0){$(1,-1),\,n=0$}}
\put(3848,1490){\makebox(0,0){$(1,-1),\,n=1$}}
\put(3855,1240){\makebox(0,0){$(-1,0),\,n=1$}}
\put(3945,1110){\makebox(0,0){$(-1,-0.5),\,n=1$}}
\put(3858,960){\makebox(0,0){$(-1,1),\,n=2$}}
\put(3808,330){\makebox(0,0){$(1,1),\,n=0$}}
\put(300,1180){%
\makebox(0,0)[b]{\shortstack{\large $a(r)$}}%
}
\put(3417,151){\makebox(0,0){6}}
\put(2948,151){\makebox(0,0){5}}
\put(2478,151){\makebox(0,0){4}}
\put(2009,151){\makebox(0,0){3}}
\put(1539,151){\makebox(0,0){2}}
\put(1070,151){\makebox(0,0){1}}
\put(600,151){\makebox(0,0){0}}
\put(540,384){\makebox(0,0)[r]{-2}}
\put(540,649){\makebox(0,0)[r]{-1}}
\put(540,915){\makebox(0,0)[r]{0}}
\put(540,1180){\makebox(0,0)[r]{1}}
\put(540,1445){\makebox(0,0)[r]{2}}
\put(540,1711){\makebox(0,0)[r]{3}}
\put(540,1976){\makebox(0,0)[r]{4}}
\end{picture}

\vspace{9mm}

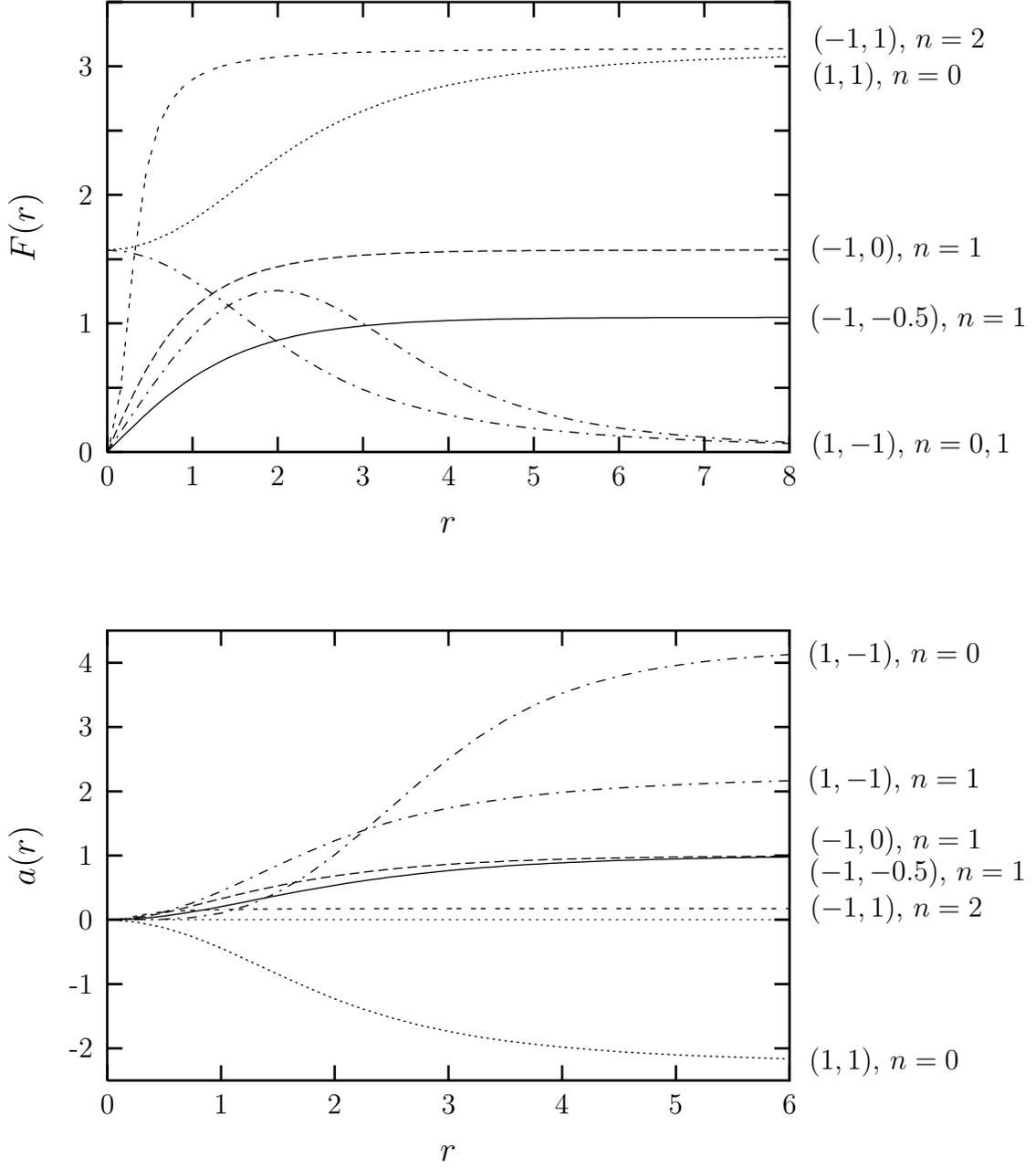
\noindent Figure 2. (a) Scalar functions $F(r)$ of various solitons and (b) 
Gauge fields $a(r)$ of various solitons: (i) Topological lump for $(-1,1)$, 
(ii) Topological vortices for $(-1,0)$ and $(-1,-0.5)$, (iii) Nontopological 
solitons for $(1,+1)$ and $(1,-1)$, (iv) Nontopological vortex for $(1,-1)$. 

\vspace{5mm}

Therefore, one can easily find all the possible rotationally symmetric soliton 
solutions by the shooting argument \cite{KKK} and they are classified in 
Table 1. Note that in Table 1 we do not include trivial solutions and the 
solutions approaching their boundary values with oscillations. Numerical 
solutions of the self-dual solitons are given in Figure 2.

In summary, we studied the Bogomol'nyi limit of a uniaxial ferromagnet system 
coupled to the Abelian CS gauge field. We found various solitons: the 
nontopological solitons and vortices, the topological vortices, and the 
topological lumps. When there is no background CS charge, there is uniquely the 
topological vortex of half winding. We make the following remarks: (i) For 
$\rho_{e}=+1(-1)$, the Bogomol'nyi equation for the nontopological solitons
and vortices reduces to the Liouville equation when $Q^{3}$ is near $-1(+1)$, 
and these solitons are the same as those in the nonrelativistic Abelian 
self-dual CS scalar model \cite{JP}. (ii) The unboundedness of the effective 
potential $V(\phi)$ in Eq. (\ref{effp}) at $\phi=\pm\infty$ disappears in the 
relativistic counterpart of our model \cite{GG} as is the case between the 
relativistic \cite{HKP} and nonrelativistic \cite{JP} Abelian self-dual CS 
scalar models. (iii) The vacuum value 
of $Q^{3}$ determined by the minimization of the anisotropy energy depends on 
the external magnetic field $H$. On the other hand, the boundary values of 
solitons $Q^{3}(\infty)$ are fixed by the background CS charge. Therefore, the 
vacuum value of $Q^{3}$ need not coincide with the boundary value of solitons 
even when the background CS charge vanishes.

We conclude with a final comment. The Bogomol'nyi limit of our model leaves no
room for free parameters. For example, the original free parameter $w$ of the
topological charge is decided by the external magnetic field $H$ and the
background charge $\rho_{e}$. Our model, therefore, possesses 
the merit of predicting the observables if the limit could be indeed realized 
in a ferromagnet sample with a given  $\lambda$.

\vspace{5mm}

We thank Q-H. Park for useful discussions. This work is supported by the 
KOSEF through the CTP at Seoul National University and the project 
number(96-0702-04-01-3, 96-1400-04-01-3), and by the Ministry of Education 
through the Research Institute for Basic Science(BSRI/96-1419, 96-2434).

\def\hebibliography#1{\begin{center}\subsection*{References
}\end{center}\list
{[\arabic{enumi}]}{\settowidth\labelwidth{[#1]}
\leftmargin\labelwidth    \advance\leftmargin\labelsep
    \usecounter{enumi}}
    \def\newblock{\hskip .11em plus .33em minus .07em}
    \sloppy\clubpenalty4000\widowpenalty4000
    \sfcode`\.=1000\relax}

\let\endhebibliography=\endlist

\begin{hebibliography}{100}
\bibitem{KIK} For a review, see A.M. Kosevich, B.A. Ivanov and A.S. Kovalev,
Phys. Rep. {\bf 194}, 117 (1990).
\bibitem{LL} For a review, see E.M. Lifshitz and L.P.P. Pitaevskii, {\it
Statistical Physics Part 2 : Landau and Lifshitz Course of Theoretical
Physics Vol. 9}, (Pergamon Press, Oxford, 1980).
\bibitem{BP} A.A. Belavin and A.M. Polyakov, JETP Lett. {\bf 22}, 245 (1975).
\bibitem{ISZ} B.A. Ivanov and V.A. Stephanovich, Phys. Lett. A {\bf 141}, 89
(1989); B.A. Ivanov, V.A. Stephanovich and A.A. Zhmudskii, J. Magn.  Magn. Mat.
{\bf 88}, 116 (1990).
\bibitem{Lee} R.A. Leese, Nucl. Phys. B {\bf 344}, 33 (1990); B {\bf 366}, 283
(1991).
\bibitem{Bog} E.B. Bogomol'nyi, Yad. Fiz. {\bf 24}, 861 (1976) [Sov. J. Nucl.
Phys. {\bf 24}, 449 (1976)].
\bibitem{HKP} J. Hong, Y. Kim and P. Y. Pac, Phys. Rev. Lett. {\bf 64}, 2230 
(1990); R. Jackiw and E. J. Weinberg, {\it ibid} {\bf 64}, 2234 (1990);
R. Jackiw, K. Lee and E. J. Weinberg, Phys. Rev. D {\bf 42}, 3488 (1990).
\bibitem{JP} R. Jackiw and S.-Y. Pi, Phys. Rev. Lett. {\bf 64}, 2969 (1990);
Phys. Rev. D {\bf 42}, 3500 (1990); {\bf 48}, 3929(E) (1993).
\bibitem{Sch} B. J. Schroers, Phys. Lett. B {\bf 356}, 291 (1995);
J. Gladikowski, B. M. A. G. Piette and B. J. Schroers, Phys. Rev. D {\bf 53},
844 (1996).
\bibitem{Nar} G. Nardelli, Phys. Rev. Lett. {\bf 73}, 2524 (1994).
\bibitem{GG} P. K. Ghosh and S. K. Ghosh, Phys. Rev. B {\bf 366}, 199 (1996);
K. Kimm, K. Lee and T. Lee, Phys. Rev. D {\bf 53}, 4436 (1996).
K. Arthur, D. H. Tchrakian and Y. Yang, Phys. Rev. D {\bf 54}, 5245 (1996).
\bibitem{op} 
P. Oh and Q-H. Park,  hep-th/9612063,  Phys. Lett. B (in press). 
\bibitem{op1} P. Oh and Q-H. Park, Phys. Lett. B {\bf 383}, 333 (1996).
\bibitem{PT} For other definition, see, {\it e.g.}, N. Papanicolaou and T.N. 
Tomaras, Nucl. Phys. B {\bf 360}, 425 (1991); E.G. Floratos, Phys. Lett. B 
{\bf 279}, 117 (1992).
\bibitem{KKK} C. Kim, S. Kim and Y. Kim, Phys. Rev. D {\bf 147}, 5434 (1993);
C. Kim and Y. Kim, Phys. Rev. D {\bf 50}, 1040 (1994).
\end{hebibliography}
\end{document}